\documentclass[prb,superscriptaddress,twocolumn,showpacs]{revtex4}

\usepackage{graphicx}
\usepackage{amsmath}
\usepackage{amsfonts}
\usepackage{bm}
\usepackage{bbm}

\usepackage{psfrag,pstricks}
\usepackage{subfigure}
\usepackage[rightcaption]{sidecap}

\begin{document}


\title{Composition dependent band gap and band edge bowing in AlInN: A combined theoretical and experimental study}

\author{S. Schulz}
\email{stefan.schulz@tyndall.ie} \affiliation{Tyndall National
Institute, Lee Maltings, Cork, Ireland}
\author{M.~A. Caro}
\affiliation{Tyndall National Institute, Lee Maltings, Cork,
Ireland} \affiliation{Department of Physics, University College
Cork, Ireland}
\author{L.-T. Tan}
\altaffiliation[Present address: ]{School of Engineering, Republic
Polytechnic, 9 Woodlands Ave 9, Singapore 738964}
\affiliation{Department of Physics, SUPA, University of Strathclyde,
Glasgow, United Kingdom}
\author{P.~J. Parbrook}
\affiliation{Tyndall National Institute, Lee Maltings, Cork,
Ireland} \affiliation{School of Engineering, University College
Cork, Ireland}
\author{R.~W. Martin}
\affiliation{Department of Physics, SUPA, University of Strathclyde,
Glasgow, United Kingdom}
\author{E.~P. O'Reilly}
\affiliation{Tyndall National Institute, Lee Maltings, Cork,
Ireland} \affiliation{Department of Physics, University College
Cork, Ireland}
%
\date{\today}

\begin{abstract}
A combined experimental and theoretical study is presented of the
band gap of AlInN, confirming the breakdown of the virtual crystal
approximation (VCA) for the conduction and valence band edges.
Composition dependent bowing parameters for these quantities are
extracted. Additionally, composition dependent band offsets for
GaN/AlInN systems are provided. We show that local strain and
built-in fields affect the band edges significantly, leading to
optical polarization switching at much lower In composition than
expected from a VCA approach.
\end{abstract}


\maketitle
The semiconductor alloy $\text{Al}_{1-x}\text{In}_x$N has a direct
band gap that spans a very wide energy range (0.69 eV to 6.25
eV).~\cite{Wu2009} This basic property makes
$\text{Al}_{1-x}\text{In}_x$N an ideal candidate for a range of
optoelectronic devices, such as laser diodes, light emitting devices
and detectors.~\cite{LiZh2012} To design
$\text{Al}_{1-x}\text{In}_x$N based devices, an accurate knowledge
of the variation of the band gap $E^\text{AlInN}_g$ with varying InN
content $x$ is required. Often the variation of the band gap of a
semiconductor alloy with composition $x$  can be successfully
described by the so-called virtual crystal approximation
(VCA)~\cite{Nord31}
\begin{equation}
E^\text{AlInN}_g=(1-x)E^\text{AlN}_g+xE^\text{InN}_g-\tilde{b}\cdot
x \cdot(1-x)\, , \label{eq:VCAEg}
\end{equation}
with a composition \emph{independent} bowing parameter $\tilde{b}$.

However, for $\text{Al}_{1-x}\text{In}_x$N systems a large range of
values for the bowing parameter $\tilde{b}$ have been reported in
the literature. Reported values scatter from \mbox{2.5 eV},
extracted from measurements on high $x$ samples, up to \mbox{10 eV}
based on measurements with low $x$ values.~\cite{KiSa97,AsDa2010}
Hence, the assumption that $\tilde{b}$ is independent of composition
has been questioned by several groups.~\cite{WaMa2008,SaBe2010}
Recently, based on density functional theory (DFT) results, the
physical mechanisms underlying this breakdown have been clarified.
It has been shown that cation-related localized states in the
conduction band (CB) and the valence band (VB) lead to the breakdown
of the VCA.~\cite{ScCa2013} These results support the assumption of
a composition \emph{dependent} bowing parameter. It is important to
note that when designing polarization matched GaN quantum wells
(QWs), using AlInN barriers,~\cite{CaSc2011,CaSc2012} the evolution
of the CB edge (CBE) and VB edge (VBE) energies with InN content $x$
is important, since this determines confinement energies for
carriers.

To shed further light on the behavior of the band gap bowing in
$\text{Al}_{1-x}\text{In}_x$N and how CBE and VBE behave with
varying InN content $x$, we have performed experimental and
theoretical studies. Our results are compared with recent
experimental literature data. We apply a tight-binding (TB) model,
to achieve an atomistic description of the electronic
structure.~\cite{CaSc2013local} This model includes local strain and
built-in fields arising from random alloy fluctuations in AlInN. The
same approach has been successfully applied to InGaN alloys
recently.~\cite{CaSc2013local}

Our results confirm that the band gap bowing parameter in
$\text{Al}_{1-x}\text{In}_{x}$N is highly composition dependent and
cannot be described by a simple VCA, Eq.~(\ref{eq:VCAEg}).
Furthermore, our calculations reveal that both CBE and VBE
separately deviate from the VCA. Therefore, we also provide
composition dependent bowing parameters for CBE and VBE in
$\text{Al}_{1-x}\text{In}_{x}$N, which can then be used as input
parameters for continuum-based descriptions, such as
$\mathbf{k}\cdot\mathbf{p}$-models, of AlInN heterostructures.
Additionally, our theoretical analysis indicates that local strain
and built-in fields, arising from random alloy fluctuations, play an
important role in the description of the band edges, contributing
therefore to the deviation from the VCA. We extract composition
dependent CB and VB offsets for GaN/$\text{Al}_{1-x}\text{In}_{x}$N
systems. Finally, we analyze the VB ordering in
$\text{Al}_{1-x}\text{In}_{x}$N. We calculate an optical
polarization switching from TM- to TE-polarized emission around
$x=0.15$.

\begin{table*}[t!]
\caption{Al$_{1-x}$In$_x$N composition dependent
bowing parameters for band gap [$b$], CB [$b^\text{CB}$], and VB
[$b^\text{VB}$], as a function of $x$.}
\begin{tabular}{l c c c c c c c c c c c c}
\hline\hline \multicolumn{1}{c}{$x$} & 0.05 & 0.08 & 0.10 & 0.13 &
0.15 & 0.18 & 0.25 & 0.35 & 0.50 & 0.65 & 0.75 & 0.85
\\\hline\smallskip
$b$ (eV) & 19.81 & 15.35 & 13.89 & 11.74 & 10.72 & 9.91 & 8.12 &
6.43 & 5.15 & 4.52 & 4.24 & 3.87
\\\smallskip
$b^\text{CB}$ (eV) & 14.01 & 10.67 & 9.29 & 7.67 & 7.01 & 6.23 &
4.95 & 3.92 & 3.08 & 2.54 & 2.22 & 1.96
\\\smallskip
$b^\text{VB}$ (eV) & -5.80 & -4.68 & -4.59 & -4.07 & -3.71 & -3.68 &
-3.17 & -2.51 & -2.07 & -1.98 & -2.01 & -1.92\\
\hline\hline
\end{tabular}
\label{tab:bowing}
\end{table*}


On the experimental side, we have grown
$\text{Al}_{1-x}\text{In}_x$N epilayers with thicknesses about 100
nm by metal organic chemical vapor deposition, using GaN nucleation
layers on $c$-plane sapphire. Here, $\text{Al}_{1-x}\text{In}_x$N
samples with $x$ ranging from 0.082 to 0.17 have been studied by
photoluminescence excitation (PLE) spectroscopy. The PLE
measurements have been performed at low temperatures  in a closed
cycle helium cryostat. The samples have been excited by a Xe-lamp
and the detected wavelength has been set to the
$\text{Al}_{1-x}\text{In}_x$N emission peak for each of the series
of samples. The PLE spectra were fitted using a sigmoidal function,
as introduced for InGaN epilayers in Ref.~\onlinecite{MaMi1999}, to
define the band gap energy together with a broadening of the
absorption edge.


On the theoretical side, band gap and band edge bowing parameters
have been studied by means of an $sp^3$ TB model. This approach
allows for a microscopic description of electronic and optical
properties of AlInN. Our TB model includes explicitly local strain
and built-in fields, arising from random alloy fluctuations. The
theoretical framework is discussed in detail in
Ref.~\onlinecite{CaSc2013local} for InGaN systems. For AlInN we used
the same approach; the required material parameters are taken from
Refs.~\onlinecite{AnORe2000,YaRi2011,CaSc2012} while the AlN TB
parameters can be found here.~\cite{footnoteTBAlN}

To gain insight into the behavior of the AlInN band gap and band
edges, we have performed TB calculations on supercells containing
approximately 12,000 atoms. The supercell is free to relax in all
three spatial directions. For each InN content $x$, calculations for
five different random configurations have been performed. The band
gaps and band edges, at each composition, have been calculated as
configurational averages. Details are given in
Ref.~\onlinecite{CaSc2013local}.

\begin{figure}[b]
\centering
\includegraphics{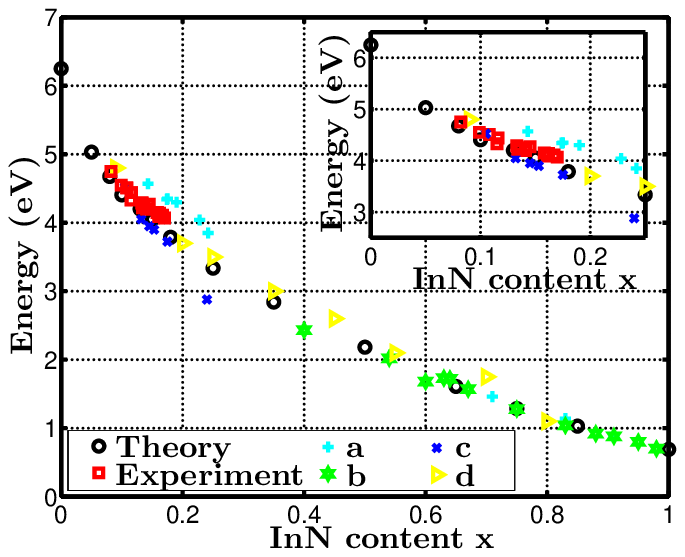}
\flushleft
\begin{tabular}{cccc}
\textbf{a:} Ref.~\onlinecite{SaBe2010}; & \textbf{b:}
Ref.~\onlinecite{JoBr2008}; & \textbf{c:}
Ref.~\onlinecite{AsDa2010}; & \textbf{d:} Ref.~\onlinecite{IlAd2008}
\end{tabular}
\caption{Band gap of $\text{Al}_{1-x}\text{In}_{x}$N as a function
of $x$. Our theoretical [open circles] and experimental results
[squares] are compared to literature data.}
\label{fig:bandgapbowing}
\end{figure}

Figure~\ref{fig:bandgapbowing} shows our theoretical TB [black
circles] and experimental PLE data [red squares] together with
literature data. Overall, we find that our TB results are in very
good agreement with the experimental data over the full composition
range. Note that the TB approach involves fitting to the band gaps
of the binary materials only. Also a closer look at the composition
range $x=0.05-0.25$, cf. inset Fig.~\ref{fig:bandgapbowing}, shows
that both our theoretical and experimental results are in good
agreement with the literature data. Furthermore, with as little as
5\% InN [$x=0.05$] in $\text{Al}_{1-x}\text{In}_x$N, a reduction in
the band gap of over 1 eV is observed. This strong reduction in band
gap can be traced back to In-related localized states in the
CB.~\cite{ScCa2013} As discussed in Ref.~\onlinecite{ScCa2013},
these localized states lead to the breakdown of the simple VCA
description. Consequently, a composition dependent bowing parameter
is required, labeled $b$ in the following.

Table~\ref{tab:bowing} summarizes the calculated values for $b$ in
$\text{Al}_{1-x}\text{In}_x$N as a function of $x$, obtained from
our TB model. This data is derived using Eq.~(\ref{eq:VCAEg}) by
fitting to the end points (binaries) and the desired $x$-value.
Especially in the low InN regime extremely large values for $b$ are
observed.

However, when modeling AlInN-based heterostructures, not only the
overall band gap bowing is important but also how the band gap
bowing is distributed between CB and VB. This quantity is of central
importance for an accurate description of electronic and optical
properties of heterostructures since it determines the confinement
energies for the carriers.

\begin{figure}[b]
\centering
\includegraphics{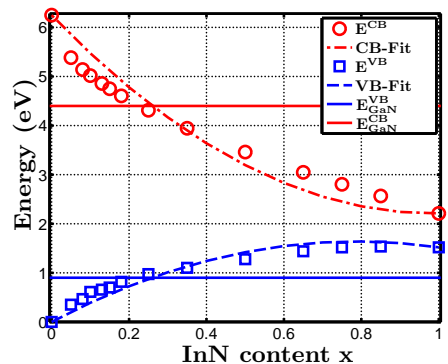}
\caption{CBE and VBE in $\text{Al}_{1-x}\text{In}_{x}$N as a
function of $x$. TB results for CBE and VBE [$E^\text{CB}$;
$E^\text{VB}$]: open symbols; dashed-(dotted) lines: fit obtained
from Eq.~(\ref{eq:VBCBbowing}). Solid horizontal lines: GaN CBE and
VBE, respectively.} \label{fig:BandEdgeBow}
\end{figure}

\begin{table*}[t!]
\caption{VB and CB offset $\Delta E^\text{GaN/AlInN}_\text{VB}$ and
$\Delta E^\text{GaN/AlInN}_\text{CB}$, respectively, between GaN and
Al$_{1-x}$In$_x$N as a function of $x$.}
\begin{tabular}{l c c c c c c c c c c c c c c}
\hline\hline \multicolumn{1}{c}{$x$} & 0 & 0.05 & 0.08 & 0.10 & 0.13
& 0.15 & 0.18 & 0.25 & 0.35 & 0.50 & 0.65 & 0.75 & 0.85 & 1
\\\hline $\Delta E^\text{GaN/AlInN}_\text{CB}$ (eV) & 1.85 & 0.98
& 0.74 & 0.62 & 0.46 & 0.35 & 0.20 & -0.09 & -0.46 & -0.94 & -1.35 &
-1.60 & -1.83 & -2.19
\smallskip\\
$\Delta E^\text{GaN/AlInN}_\text{VB}$ (eV) & 0.90 & 0.55 & 0.43 &
0.29 & 0.24 & 0.20 & 0.08 & -0.07 & -0.20 & -0.38 & -0.54 & -0.62 &
-0.64 & -0.62\\
\hline\hline
\end{tabular}
\label{tab:deltaE}
\end{table*}

Figure~\ref{fig:BandEdgeBow} shows the TB results for the CBE
[circles] and VBE [squares] in $\text{Al}_{1-x}\text{In}_x$N as a
function of $x$. A VCA fit to the TB data over the whole composition
range is given by the dashed-(dotted) lines and is obtained from
\begin{eqnarray}
\nonumber E^\text{AlInN}_\text{CB}&=& x(E^\text{InN}_g+\Delta
E_\text{VB})+(1-x)E^\text{AlN}_g-\tilde{b}^\text{CB}x(1-x)\, ,\\
E^\text{AlInN}_\text{VB}&=&x\Delta E_\text{VB} -\tilde{b}^\text{VB}x
(1-x)\,\, . \label{eq:VBCBbowing}
\end{eqnarray}
Here, $\tilde{b}^\text{CB}$ and $\tilde{b}^\text{VB}$ are
composition \emph{independent} and the VB offset $\Delta
E_\text{VB}$ between InN and AlN is taken from
Ref.~\onlinecite{KiVe2007}. Figure~\ref{fig:BandEdgeBow} confirms
again that composition \emph{independent} bowing parameters fail to
describe the CBE and VBE in AlInN. The behavior of the CBE can
clearly not be described by the VCA [dashed-dotted line]. The VBE
shows a similar behavior, but with a smaller deviation from a
VCA-like model. To a first approximation a composition
\emph{independent} VB bowing parameter could be used
[$\tilde{b}^\text{VB}=-2.64$ eV].\\ We apply the procedure described
above for the composition dependent band gap bowing parameter $b$ to
CBE and VBE. The values for CBE and VBE bowing parameters
$b^\text{CB}$ and $b^\text{VB}$, respectively, are summarized in
Table~\ref{tab:bowing}. From Table~\ref{tab:bowing} we conclude that
the very strong composition dependence of the AlInN band gap in the
low InN regime mainly arises from the composition dependence of the
CBE. In this regime the CBE bowing parameter $b^\text{CB}$ is much
larger than the VBE parameter $b^\text{VB}$. This finding ties in
with recent DFT results on the low InN regime.~\cite{ScCa2013}
However, since we observe also a significant VBE bowing parameter,
the commonly applied assumption~\cite{AmMa2002} in which all the
bowing is attributed to the CB seems to fail in AlInN, especially
when studying higher InN contents ($x>0.5$), since $b^\text{CB}$ and
$b^\text{VB}$ are comparable in magnitude for that range.

\begin{figure}[b]
\centering
\includegraphics{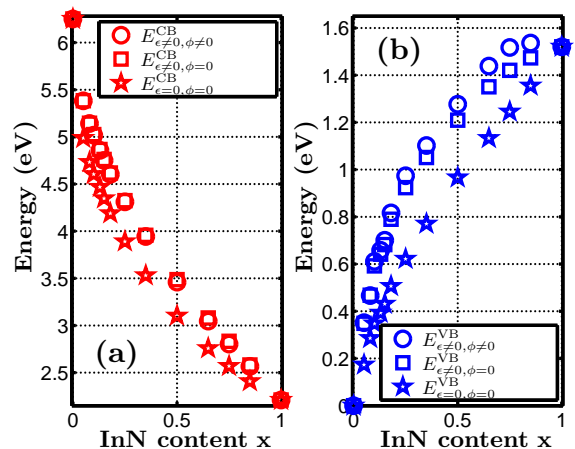}
\caption{CBE (a) and VBE (b) of Al$_{1-x}$In$_x$N as a function of
$x$. Open stars: Without strain and built-in potential
($E^\text{VB/CB}_{\epsilon=0,\phi=0}$); Open squares: With strain
but without built-in potential
($E^\text{VB/CB}_{\epsilon\neq0,\phi=0}$); Open circles: With strain
and built-in potential
($E^\text{VB/CB}_{\epsilon\neq0,\phi\neq0}$).}
\label{fig:LocalEffectsCBVB}
\end{figure}

Having discussed the composition dependence of both the band gap and
band edges in $\text{Al}_{1-x}\text{In}_x$N, we turn now and focus
on the effect that local alloy, strain and built-in field
fluctuations have on the results. In Fig.~\ref{fig:LocalEffectsCBVB}
we disentangle the impact of these quantities on CBE
[Fig.~\ref{fig:LocalEffectsCBVB} (a)] and VBE
[Fig.~\ref{fig:LocalEffectsCBVB} (b)]. We start with the analysis of
the CBE. Comparing the results in the absence of the local strain
and built-in potential contributions
[$E^\text{CB}_{\epsilon=0,\phi=0}$; open stars] with data when
including local strain effects only
[$E^\text{CB}_{\epsilon\neq0,\phi=0}$; open squares] we observe a
strong shift of the CBE to higher energies. This shift to higher
energies arises from the deformation potential correction due to
local hydrostatic strain, since the interatomic bond lengths of InN
are larger ($\simeq 14$\%) than those of AlN. When also including
the local built-in potential fluctuations
[$E^\text{CB}_{\epsilon\neq0,\phi\neq0}$; open circles], we observe
almost no difference between this full calculation and the situation
where we have switched off the built-in potential fluctuations
[$E^\text{CB}_{\epsilon\neq0,\phi=0}$; open squares], at least on
the energy scale shown.

Figure~\ref{fig:LocalEffectsCBVB} (b) shows the situation for the
VBE. Taking local strain effects into account but neglecting local
built-in potential contributions
[$E^\text{VB}_{\epsilon\neq0,\phi=0}$; open squares], one observes a
shift of the VBE to higher energies compared to the situation when
only local alloy effects [$E^\text{VB}_{\epsilon=0,\phi=0}$; open
stars] are included. The effect is mainly related to local
compressive strains. In the presence of the local built-in field
contributions [$E^\text{VB}_{\epsilon\neq0,\phi\neq0}$; circles], we
observe a further upward bowing of the VBE compared to the situation
without the local built-in potential
[$E^\text{VB}_{\epsilon\neq0,\phi=0}$, open squares]. These results
are similar to our findings on the band edges in
$\text{In}_x\text{Ga}_{1-x}$N.~\cite{CaSc2013local}

Having discussed how band edges change in
$\text{Al}_{1-x}\text{In}_x$N with $x$, we estimate in the following
the composition dependent CB and VB offsets $\Delta
E^\text{GaN/AlInN}_\text{CB}$ and $\Delta
E^\text{GaN/AlInN}_\text{VB}$, respectively, between GaN and AlInN.
Here, we estimate band offsets in the absence of strain and
polarization fields. When band offsets are included, for example, in
QW calculations by means of $\mathbf{k}\cdot\mathbf{p}$-theory,
strain and built-in potentials will be added
separately.~\cite{VoWibook} Here, the VB offset \mbox{$\Delta
E^\text{GaN/AlInN}_\text{VB}$} is calculated as \mbox{$\Delta
E^\text{GaN/AlInN}_\text{VB}=\Delta
E^\text{GaN/AlN}_\text{VB}-E^\text{AlInN}_\text{VB}$}, where
$E^\text{AlInN}_\text{VB}$ is obtained from
Eqs.~(\ref{eq:VBCBbowing}) using data from Table~\ref{tab:bowing}.
\mbox{$\Delta E^\text{GaN/AlN}_\text{VB}$} denotes the VB offset
between GaN and AlN. We assume \mbox{$\Delta
E^\text{GaN/AlN}_\text{VB}=0.9$ eV}, which is in the range of
reported literature values [$\Delta
E^\text{GaN/AlN}_\text{VB}=0.15-1.4$
eV],~\cite{MaBo96,Moen96,BiFe2001,MoMi2011} and at $x=1$,
\mbox{$|E^\text{GaN/InN}_\text{VB}|=0.62$ eV} in accordance with
Ref.~\onlinecite{MoMi2011} for the VB offset between InN and GaN.
Our approach is similar to the approach used in
Ref.~\onlinecite{AmMa2002}. Therefore, if $\Delta
E^\text{GaN/AlInN}_\text{VB}>0$ the VBE in GaN is at higher energies
than the VBE in AlInN. The composition dependent CB offset, $\Delta
E^\text{GaN/AlInN}_\text{CB}$, is calculated as \mbox{$\Delta
E^\text{GaN/AlInN}_\text{CB}=E^\text{AlInN}_\text{CB}-(\Delta
E^\text{GaN/AlN}_\text{VB}+E^\text{GaN}_g)$}, where
\mbox{$E_g^\text{GaN}$} is the bulk band gap of GaN.~\cite{Wu2009}
$E^\text{AlInN}_\text{CB}$ is calculated from
Eqs.~(\ref{eq:VBCBbowing}) using data from Table~\ref{tab:bowing}.
Here, $\Delta E^\text{GaN/AlInN}_\text{CB}>0$ indicates that the CBE
in GaN is at lower energies than the CBE in AlInN.

The obtained results are summarized in Table~\ref{tab:deltaE} where
we estimate that the VB and CB offsets are positive up to 25\% InN
in AlInN. In terms of a heterostructure, neglecting strain and
built-in potentials, this indicates that electrons and hole are
confined in the GaN region. For $x\geq0.25$ we observe a change in
sign in CB and VB offsets, indicating that the carriers are confined
in the AlInN region. We note that there is an uncertainty in the
calculated composition values where CB and VB offset change sign due
to the uncertainty in the AlN/GaN VB
offset;~\cite{MaBo96,Moen96,BiFe2001,MoMi2011} an increase
(decrease) in the assumed VB offset would for instance lead to the
CB crossover occurring at lower (higher) InN compositions.

\begin{figure}[t!]
\centering
\includegraphics{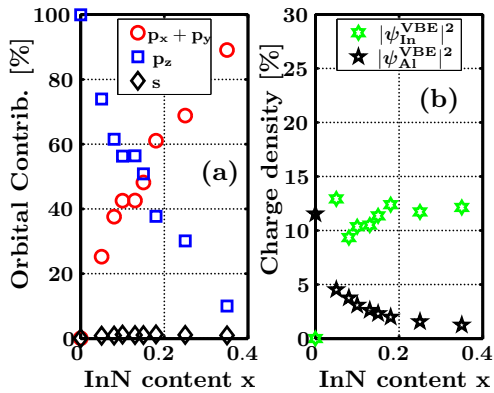}
\caption{(a) Orbital contributions to the VBE. (b) Charge density at
In- and Al-atoms for the VBE. The data is shown as a function of $x$
in $\text{Al}_{1-x}\text{In}_x$N.} \label{fig:OribtContri}
\end{figure}


In the final step we study how the optical polarization due to the
VB ordering changes in $\text{Al}_{1-x}\text{In}_x$N with $x$.
Compared to more conventional III-V semiconductors such as InAs, the
spin-orbit coupling (SOC) in the group-III nitrides is
small.~\cite{VuMe2001,VuMe2003} Neglecting the weak SOC, the topmost
VB in AlN is $p_z$-like while in InN it is a linear combination of
$p_x$- and $p_y$-like states. Figure~\ref{fig:OribtContri} (a) shows
the contribution of the $s$-, $p_x$-, $p_y$- and $p_z$-like orbitals
to the VBE in fully relaxed $\text{Al}_{1-x}\text{In}_x$N as a
function of $x$. Our TB results show that below $x=0.15$, the
dominant orbital contribution is still $p_z$-like, while for
$x>0.15$ the linear combination of $p_x$- and $p_y$-like states
starts to dominate. Therefore, our data indicates an optical
polarization switching [TM to TE] in $\text{Al}_{1-x}\text{In}_x$N
at $x=0.15$. Performing calculations for AlInN systems
pseudomorphically grown on GaN, the polarization switching occurs at
$x\approx 0.18$ (not shown). Figure~\ref{fig:OribtContri} (b)
illustrates that the charge density on the In-sites
($|\psi^\text{VBE}_\text{In}|^2$) increases significantly compared
to the Al-sites ($|\psi^\text{VBE}_\text{Al}|^2$) with as little as
5\% InN in the system (remainder of the charge density is on the
N-sites). This indicates a strong localization of the wave function
around the In sites, in agreement with recent DFT
data,~\cite{ScCa2013} and explains the surprisingly early onset of
the polarization switching.


In summary we have studied the band gap bowing of AlInN as a
function of the InN content $x$ both experimentally and
theoretically. Our atomistic TB results are in good agreement with
the performed PLE measurements and with experimental literature
data. We confirm that the assumption of a composition
\emph{independent} bowing parameter fails and provide data for the
composition dependence of the bowing parameter. Moreover, we find
that both CBE and VBE show deviations from a simple VCA description.
Composition dependent VBE and CBE
bowing parameters have been extracted.\\
Our microscopic analysis reveals that local strain and built-in
field effects play a significant role in the composition dependent
behavior of CBE and VBE. We have used this data to study the band
offsets in AlInN/GaN systems. Our analysis of the optical
polarization in AlInN shows a switching from TM- to TE-polarized
emission at $15-18$\% InN.

This work was supported by Science Foundation Ireland (project No.
10/IN.1/I2994, 10/IN.1/I2993 and 07/EN/E001A), the Engineering and
Physical Sciences Research Council and the European Union Project
ALIGHT (FP7-280587). We are grateful to the crystal growth teams at
Strathclyde University, Cambridge University and CRHEA-CNRS,
Valbonne who provided the AlInN samples measured for this work.





\end{document}